\documentclass[sigconf,screen]{acmart}
\AtBeginDocument{%
  }

\acmYear{2026}\copyrightyear{2026}
\setcopyright{cc}
\setcctype[4.0]{by-nc-nd}
\acmConference[FSE Companion '26]{34th ACM Joint European Software Engineering Conference and Symposium on the Foundations of Software Engineering}{July 5--9, 2026}{Montreal, QC, Canada}
\acmBooktitle{34th ACM Joint European Software Engineering Conference and Symposium on the Foundations of Software Engineering (FSE Companion '26), July 5--9, 2026, Montreal, QC, Canada}
\acmDOI{10.1145/3803437.3806414}
\acmISBN{979-8-4007-2636-1/26/07}

\usepackage{xr}
\usepackage{soul}
\usepackage{makecell}
\usepackage{booktabs}
\usepackage{enumitem}
\usepackage{multirow}
\usepackage{xcolor}
\usepackage{xspace}
\usepackage[normalem]{ulem} 
\usepackage{listings}
\usepackage[htt]{hyphenat}
\usepackage{tikz}

\definecolor{LightGray}{gray}{0.95}

\definecolor{darkgreen}{HTML}{59B02E}
\definecolor{darkred}{HTML}{B87213}
\definecolor{lightgray}{rgb}{0.97,0.97,0.97}
\definecolor{LightGray}{gray}{0.9}

\newcommand{\drawcomments}{}

\newcommand*\circled[1]{\tikz[baseline=(char.base)]{
            \node[shape=circle,draw,inner sep=0.5pt] (char) {#1};}}

\providecommand{\st}[1]{#1}

\ifdefined\drawcomments

\newcommand{\byron}[1]{\textcolor{purple}{[Byron: #1]}}
\newcommand{\lili}[1]{\textcolor{cyan}{[Lili: #1]}}

\newcommand{\solved}[1]{\textcolor{red}{\textbf{Solved:} \{ #1 \}}}

\newcommand{\revision}[1]{\textcolor{blue}{#1}}
\newcommand{\delete}[1]{\textcolor{red}{\st{#1}}}
\newcommand{\newcontent}[1]{\textcolor{orange}{#1}}
\newcommand{\sub}[2]{\delete{#1}\ \newcontent{#2}}
\newcommand{\todo}[1]{\textcolor{green}{TODO: #1}}

\newcommand{\agent}{MIMIC\xspace}
\newcommand{\agentPy}{MIMIC-Py\xspace}

\newcommand{\ODY}{ODYSSEY\xspace}

\else

\DeclareRobustCommand{\byron}[1]{}
\DeclareRobustCommand{\lili}[1]{}
\DeclareRobustCommand{\revision}[1]{}
\DeclareRobustCommand{\solved}[1]{}
\DeclareRobustCommand{\delete}[1]{}
\DeclareRobustCommand{\newcontent}[1]{}
\DeclareRobustCommand{\sub}[2]{}
\DeclareRobustCommand{\todo}[1]{}

\fi

\begin{document}

\title{\agentPy: An Extensible Tool for Personality-Driven Automated Game Testing with Large Language Models}

\author{Yifei Chen}
\email{yifei.chen@mail.mcgill.ca}
\orcid{0009-0006-8638-0818}
\affiliation{%
  \institution{Electrical and Computer Engineering, McGill University}
  \city{Montréal}
  \state{Québec}
  \country{Canada}
}

\author{Sarra Habchi}
\email{sarra.habchi@cohere.com}
\orcid{0000-0002-5989-1413}
\affiliation{%
  \institution{Cohere}
  \city{Montréal}
  \state{Québec}
  \country{Canada}
}

\author{Lili Wei}
\authornote{Lili Wei is the corresponding author.}
\email{lili.wei@mcgill.ca}
\orcid{0000-0002-2428-4111}
\affiliation{%
  \institution{Electrical and Computer Engineering, McGill University}
  \city{Montréal}
  \state{Québec}
  \country{Canada}
}



\begin{abstract}

Modern video games are complex, non-deterministic systems that are difficult to test automatically at scale. Although prior work shows that personality-driven Large Language Model (LLM) agents can improve behavioural diversity and test coverage, existing tools largely remain research prototypes and lack cross-game reusability.

This tool paper presents \agentPy, a Python-based automated game-testing tool that transforms personality-driven LLM agents into a reusable and extensible framework. \agentPy exposes personality traits as configurable inputs and adopts a modular architecture that decouples planning, execution, and memory from game-specific logic. It supports multiple interaction mechanisms, enabling agents to interact with games via exposed APIs or synthesized code. We describe the design of \agentPy and show how it enables deployment to new game environments with minimal engineering effort, bridging the gap between research prototypes and practical automated game testing.

The source code and a demo video are available on our project webpage: \href{https://mimic-persona.github.io/MIMIC-Py-Home-Page/}{https://mimic-persona.github.io/MIMIC-Py-Home-Page/}.

\end{abstract}



\received{22 January 2026}

\maketitle

\section{Introduction}
\label{sec:introduction}

\textbf{Motivation.} Modern video games have become one of the most significant sectors of the entertainment industry~\cite{Newzoo_2026}, making the maintenance of software quality through rigorous testing increasingly critical~\cite{GameTestingStats}. However, the complexity of modern games, characterized by their rich state spaces and non-deterministic environments, poses substantial challenges for automated testing techniques. As a result, most game studios still rely heavily on manual testing, which is costly, time-consuming, and difficult to scale.

\textbf{Related Work.} 
Prior work has explored agent-based game testing using Machine Learning (ML) techniques such as Reinforcement Learning (RL) and Imitation Learning (IL) to reduce human effort~\cite{RLTesting1, ILTesting2}. However, RL-based methods depend on well-engineered reward functions and IL-based methods rely on expert demonstrations, making both require substantial manual effort to adapt to new games, tasks, or evolving game versions~\cite{61AutoMC2024}, limiting their practicality as testing tools.

More recently, Large Language Models (LLMs) have been applied to game-playing agents and demonstrated stronger adaptability with less human effort across diverse game environments~\cite{OpenAI5DefeatsDota2Team, Voyager2023, GITM2023, ChessGPT2023, startCraftRL2019, BugCraft_ASE2025}. Despite these advances, relatively little work has explored the use of LLM-based agents as practical game-testing tools, particularly with respect to deployability and extensibility.

A further limitation shared by both ML-based agents and LLM-based agents is that they overlook behavioural diversity. Human players often adopt different strategies for similar tasks, shaped by individual personality traits~\cite{AggressivePlayerBehaviour2008}. However, most existing agents exhibit homogeneous, repetitive behaviour, limiting their ability to thoroughly explore game states and reducing the effectiveness of automated testing for complex game environments.

\textbf{Our Contribution.}
To address these challenges, we previously proposed \agent~\cite{MIMIC_YIFEI_ASE_2025}, an LLM-based framework that integrates gameplay personality traits to generate diverse solutions for similar in-game tasks and achieve broader coverage. This design is grounded in empirical studies showing strong correlations between player personality traits and in-game behaviour~\cite{WCPersonalityBehaviour2011, WCPersonalityBehaviour2014}.
By embedding personality-driven decision-making into game-playing agents, \agent systematically explores diverse gameplay strategies in the same game situation.
For example, when encountering an opponent, agents with different personality traits may respond differently. A cautious agent may choose to escape to minimize risk; an aggressive agent may engage directly to defeat the opponent as quickly as possible; while a more adrenaline-seeking agent may prefer a riskier, more challenging way to win.
This difference is important for testing as it enables the exploration of a wider range of gameplay behaviours and game states rather than repeatedly following a single dominant strategy.

In our prior evaluation, \agent demonstrated effectiveness across multiple games of varying scale, including \emph{Dungeon Adventures}\cite{DA}, \emph{Shattered Pixel Dungeon}\cite{SPD}, and \emph{Minecraft}\cite{Minecraft}. In larger-scale settings, \agent consistently outperformed random-based baselines by up to \textbf{1.30$\times$} in branch coverage and \textbf{14.46$\times$} in interaction-level coverage. When evaluated in \emph{Minecraft} against the state-of-the-art agent, \ODY~\cite{56Odyssey2024}, \agent solved more complex, multi-step tasks, while exhibiting substantially greater behavioural diversity. 
These results suggest that integrating personality-driven behaviours enhances both problem-solving capability and exploration effectiveness in automated game testing. Full experimental details are reported in our prior work~\cite{MIMIC_YIFEI_ASE_2025} and partial evaluation results are available on our project webpage~\cite{MIMIC-Py_Webpage}.

\textbf{Novelty of This Tool Demonstration Paper.}
Porting automated test-generation tools to new games remains challenging due to heterogeneous game architectures, diverse interaction interfaces, and the lack of standardized testing APIs~\cite{SurveyVideoGameTesting}. As a result, most existing game testing tools are tightly coupled to a single game or engine, limiting their reuse and practical adoption.

Building on the original \agent framework, this paper presents \agentPy, a Python-based implementation designed to be a reusable, extensible testing tool. 
The novelty of this tool paper lies in re-engineering the original JavaScript research prototype into a more practical and customizable system for deployment. In particular, compared with the original JavaScript prototype, \agentPy\ exposes more reusable integration interfaces and provides more centralized configuration support, making it easier to adapt the system to new games and customize experimental settings. While \agentPy\ improves usability and extensibility, it does not introduce new agentic components beyond those in the original prototype.

Importantly, \agentPy preserves the original framework’s ability to interact with games via both structured action plans and executable behaviors, while re-engineering these mechanisms as configurable and extensible modules suitable for practical use. By decoupling planning, action execution, and self-summarization, \agentPy significantly reduces the engineering effort required to port agents to new game environments.

This paper focuses on \agentPy as a practical testing tool, detailing its workflow, extensibility mechanisms, and deployment process. Detailed algorithmic design and comprehensive evaluation results are available in our prior work~\cite{MIMIC_YIFEI_ASE_2025} and are therefore not repeated here. 
A demo video illustrating how to run \agentPy across games is available at: \href{https://youtu.be/qc2kvDUmgxk}{https://youtu.be/qc2kvDUmgxk}.

The paper is organized as follows: Section~\ref{sec:implementation} presents the design and implementation of \agentPy as a reusable automated game testing tool. Section~\ref{sec:extensibility} describes how \agentPy can be extended and deployed to new game environments. Finally, Section~\ref{sec:conclusion} concludes the paper and outlines future directions.

\section{\agentPy}
\label{sec:implementation}

\begin{figure}[!t]
    \centering
    \includegraphics[width=\linewidth]{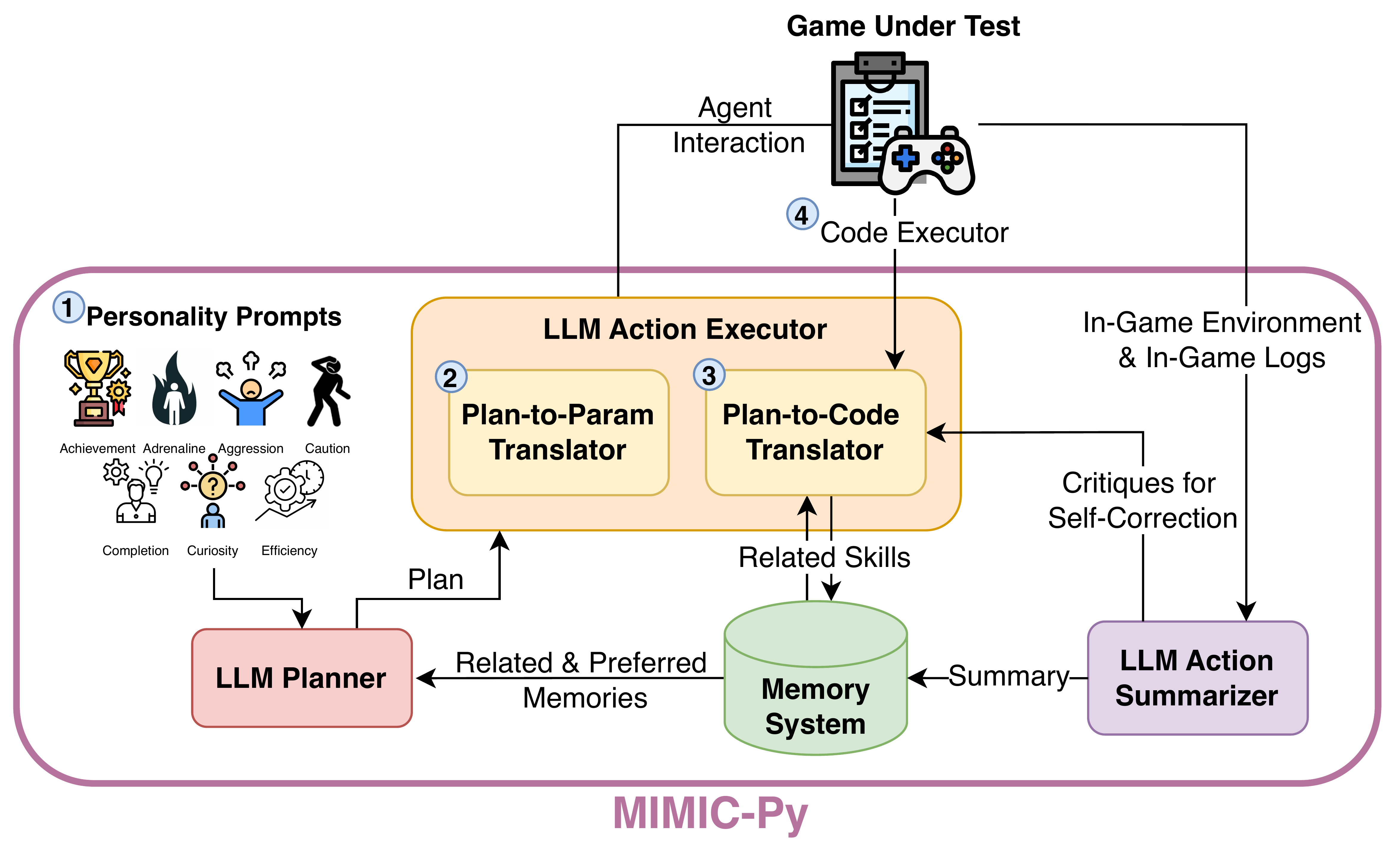}
    \caption{Overview of the \agentPy framework.}
    \label{fig:overview}
\end{figure}

\figurename~\ref{fig:overview} presents an overview of \agentPy, a personality-driven agent-based game testing tool designed for diverse gameplay behaviors, scalable testing, and lightweight adaptation to new game environments. The system consists of four core components: the \emph{Planner}, \emph{Action Executor}, \emph{Action Summarizer}, and \emph{Memory System}.

At runtime, \agentPy operates in an iterative loop. Given a testing objective and a personality trait, the Planner generates an action plan, which the Action Executor translates into concrete interactions with the game under test. The Action Summarizer evaluates the execution outcome and records a structured summary in the Memory System. Retrieved memories then inform subsequent planning, enabling long-running and context-aware testing.

\subsection{Planner}
\label{subsec:planner}


The LLM Planner generates action plans during testing while incorporating personality traits to explore diverse strategies for similar testing tasks, enabling varied interaction traces without manually defining multiple behaviours.

To ensure \agentPy can be applied to games with varying task complexity, the Planner integrates a Hybrid Planning strategy that combines Bottom-Up and Top-Down Planning. Bottom-Up Planning supports reactive, fine-grained interactions, which are effective for short-horizon or exploratory behaviours. Top-Down Planning decomposes high-level objectives into sub-tasks, supporting long-horizon reasoning and goal tracking. This Hybrid design allows \agentPy to operate robustly across both exploratory and multi-step game environments without environment-specific tuning.

The Planner also conditions its decisions on the current game state and relevant past experiences retrieved from the Memory System (Section~\ref{subsec:memory-system}), helping maintain behavioural consistency and avoid redundant or ineffective actions across iterations.

\subsubsection{\textbf{Personality Model and Configuration}}

\agentPy takes the PathOS personality model~\cite{ArtificialPlayers2020} to support personality-driven testing behaviours. PathOS defines seven behaviourally grounded personality traits, \textit{Achievement}, \textit{Adrenaline}, \textit{Aggression}, \textit{Caution}, \textit{Completion}, \textit{Curiosity}, and \textit{Efficiency}, synthesized from multiple player modelling studies. These traits capture common gameplay preferences and strategies, making them well-suited for driving diverse in-game behaviours.


Personality traits are injected into the Planner as configurable text prompts rather than hard-coded behaviours, allowing users to select or define personalities without changing the planning logic.

To support deployment across game environments, \agentPy defines lightweight mappings from game entity types predefined by PathOS to semantically equivalent concepts in each target game. PathOS defines nine entity types in total. For example, \emph{Enemy Hazard}, described as “A hostile character, etc., which could incite combat”, can be mapped to “enemies” in some games, and to “mobs” in others.
These mappings enable personality prompts to be reused across games with minimal configuration changes. 
The full list of entities can be found on our project website~\cite{MIMIC-Py_Webpage}.

\subsubsection{\textbf{Hybrid Planning}}
Many existing LLM-based agents generate only the next immediate action from the current game state~\cite{Voyager2023, zhao2024thinkembodiedagentvirtual, ALFWorld20}, a strategy we refer to as \emph{Bottom-Up Planning}. While effective for short-horizon or exploratory tasks, this approach often struggles with complex testing objectives that require long-horizon reasoning and sustained goal tracking~\cite{GITM2023}. For example, when tasked with crafting an in-game tool, an agent may successfully collect the required resources but later divert them to unrelated actions, ultimately failing to complete the original objective.

To address this limitation, \agentPy employs a \textbf{Hybrid Planner} that dynamically switches between Bottom-Up and Top-Down strategies to track goals and task progress better. Bottom-Up Planning enables fine-grained, reactive interactions, while Top-Down Planning decomposes high-level objectives into ordered sub-tasks, providing explicit goal structure for long-horizon execution. By integrating both modes, \agentPy supports complex testing objectives while retaining flexibility for exploratory behaviours, without requiring environment-specific planning logic.

To improve robustness, the Hybrid Planner further incorporates mechanisms for plan validation and revision that detect and correct infeasible or inconsistent plans, ensuring alignment with game rules and available interactions. Additional algorithmic details are provided in our prior work~\cite{MIMIC_YIFEI_ASE_2025}.

\subsection{Action Summarizer}
\label{subsec:summarizer}

The LLM Action Summarizer evaluates each executed plan and produces structured summaries of the interaction outcome, providing reusable and interpretable feedback to the Planner. 
These summaries, including the outcomes of planned actions and their relevant context, are stored in the Memory System and retrieved in subsequent iterations to guide future planning (Section~\ref{subsec:memory-system}). This process allows \agentPy\ to adapt over time while remaining aligned with the selected personality trait.

\subsection{Action Executor}
\label{subsec:executor}

The Action Executor bridges \agentPy\ to the game under test by translating action plans into executable interactions. To support diverse game control interfaces with minimal engineering effort, it provides two mechanisms: \emph{Plan-to-Parameters} and \emph{Plan-to-Code}.

\subsubsection{\textbf{Plan-to-Parameters Translator}}

When a game exposes well-defined APIs that directly map to all in-game actions, the Action Executor translates high-level plans into API input parameters, enabling efficient interaction without code generation and supporting environments with mature control interfaces.

\subsubsection{\textbf{Plan-to-Code Translator}}

Some games expose only low-level APIs or SDKs that support basic actions, requiring testers to manually assemble code scripts for more complex behaviours. This is the case for \emph{Minecraft}~\cite{Minecraft}, where interaction relies on the Mineflayer API~\cite{Mineflayer}, offering primitive controls but limited support for advanced actions~\cite{MIMIC_YIFEI_ASE_2025}. 

To operate in such environments, the Plan-to-Code Translator converts high-level plans into executable code snippets that invoke available APIs. These generated scripts, referred to as \emph{Skills}, encapsulate reusable interaction logic that \agentPy can invoke directly in subsequent executions. 
When a Skill fails to achieve its intended plan, the Action Summarizer provides feedback to guide the Translator's iterative refinement.

This iterative Skill construction design, centered on a growing Skill Library,
enables \agentPy to interact with games that lack comprehensive testing APIs without extensive manual scripting. Only a small set of example API usages is required as initial basic Skills, which guide the Translator to compose valid calls and serve as reusable functions for generating more advanced Skills. As a result, \agentPy remains lightweight and practical to deploy in new or evolving game environments.

\subsubsection{\textbf{Custom Translators}}

Beyond the built-in translators, the Action Executor supports customization for games with unique or non-standard interaction mechanisms.
A custom translator specifies how Planner outputs are interpreted and mapped to game-specific actions or execution routines, and serves as a middleware layer that reuses \agentPy's existing communication and feedback infrastructure to connect Planner outputs to game-side execution.

Implementing a custom translator requires only:
(1) defining the Planner’s expected output format via prompt configuration;
(2) receiving plans through the existing socket-based interface; 
(3) executing the corresponding game actions using the game’s APIs or SDKs; and
(4) returning structured execution feedback (e.g., observations, logs, or errors) through the same channel.
No changes to the Planner, Memory System, or Action Summarizer are required.

This extensibility is enabled by the Action Executor's encapsulated design, which isolates all game-facing logic behind a small set of interaction APIs.

The Translator type (Plan-to-Parameters, Plan-to-Code, or Custom) is selected via configuration file and determines how Planner outputs are interpreted and executed, as detailed in Section~\ref{subsec:deploy-new-game}.

\subsection{Memory System}
\label{subsec:memory-system}

The Memory System stores past interactions, including executed actions, environmental contexts, and execution summaries, collectively referred to as \emph{Memories}, as well as reusable interaction code (\emph{Skills}). These records are retrieved during planning and code synthesis to support context-aware decision-making and maintain personality-consistent behaviour across testing iterations.

To support long-running sessions, the Memory System retrieves only the most relevant Memories and Skills at each iteration, rather than injecting the full interaction history into each LLM prompt. This design avoids input-length constraints, reduces inference imprecision, and enables effective reuse of prior experience.

\agentPy adopts a Retrieval-Augmented Generation (RAG) approach~\cite{11RAGOriginalPaper} to achieve this. Both Memories and Skills are embedded in a vector database (ChromaDB~\cite{chromaDB}) for similarity-based retrieval during planning and execution. This mechanism reduces token overhead and improves inference robustness~\cite{RAGSurveyGAo2024, YuHaoRAGEvaluation2024}.

The Memory System retrieves three types of information: (1) \emph{\textbf{Preferred Memories}} aligned with the selected personality trait; (2) \emph{\textbf{Related Memories}} from similar past game states; and (3) \emph{\textbf{Related Skills}} implementing interaction logic relevant to the current plan.

For personality alignment, each Memory is augmented with an LLM-generated preference summary describing how the action and its outcome reflect a given personality trait. For situational relevance, Memories store the game state at execution time. Skills are stored with natural-language descriptions of their functionality.

At runtime, retrieval uses cosine similarity over vector embeddings. Preferred Memories are retrieved by matching preference summaries against the active personality prompt, while related Memories are retrieved by matching the current game state against the stored one. Skills are retrieved by matching the current plan description against Skill descriptions. In all cases, only the top-$k$ entries ($k=5$ by default) are passed to the Planner or Action Executor, enabling efficient reuse of prior experience while keeping each decision step focused and computationally efficient.

\section{Extensibility and Deployment}
\label{sec:extensibility}

This section describes how users extend and deploy \agentPy in practice. We focus on the concrete extension points exposed by the tool and the effort required to adapt it to new testing scenarios. 

\subsection{Extending Personality Profiles}
\label{subsec:extend-personality}

To support diverse testing behaviours, \agentPy encodes personality profiles as textual prompts provided directly to the Planner (\circled{1} in \figurename~\ref{fig:overview}), decoupling behavioural variation from the underlying implementation. Users can extend the personality set by adding new prompts and selecting them at runtime.

\subsection{Deploying into New Game Environments}
\label{subsec:deploy-new-game}

Deploying \agentPy to a new game environment centers on extending the Action Executor, which translates the Planner's outputs into concrete game interactions. Depending on the interfaces exposed by the target game, users can extend the plan-to-action translators in one of two ways (Section~\ref{subsec:executor}). Both options require only localized, game-oriented changes.

\paragraph{\textbf{Option 1: API-Driven and Custom Interaction Translators}}

\agentPy provides a unified, socket-based interaction mechanism for translating Planner outputs into game actions, which supports both the built-in \emph{Plan-to-Parameters Translator} and fully custom translators.
In this pathway, the Planner produces structured plans in a user-defined format and sends them to the game environment via the built-in WebSocket interface. \agentPy then blocks until execution feedback is returned on the same channel.

When a game exposes well-defined action APIs, users can directly adopt the Plan-to-Parameters Translator (\circled{2} in \figurename~\ref{fig:overview}) by adapting the Planner prompt templates so that generated plans conform to the parameter structure expected by the game's APIs.
For games with non-standard or proprietary interaction mechanisms, users may instead define custom plan formats and implement corresponding \emph{environment-side handlers} that interpret and execute these plans within the game runtime, while reusing \agentPy's existing communication and feedback protocol.

\paragraph{\textbf{Option 2: Code-Centric Interaction}}
For games that provide only low-level APIs or SDKs, users can enable the built-in \emph{Plan-to-Code Translator} (\circled{3} in \figurename~\ref{fig:overview}). In this mode, the Translator generates executable code snippets for interaction. 

To bootstrap this process, users need to provide:
(1) game specifications that include example code snippets and textual descriptions, 
and
(2) a small set of initial Skills (helper functions) built on top of these APIs.
These examples guide the code-generation process and enable the incremental construction of more advanced Skills.

In addition, users need to implement a \emph{code executor} (\circled{4} in \figurename~\ref{fig:overview}), which is a game-specific runtime component responsible for executing code generated by the Plan-to-Code Translator inside the game environment and returning execution feedback (e.g., observations, logs, and errors) to \agentPy. The executor is invoked by the Action Executor at runtime and serves as the integration point for Plan-to-Code interaction in a new game.


Across both options, the interaction pathway is selected through configuration, while all game-specific implementation is confined to the Action Executor and the corresponding \emph{environment-side handlers}, which execute action plans or code and return feedback from the game runtime. 
As a result, adapting \agentPy\ to a new environment mainly requires updating the game-specific information and examples used in seven prompts, together with minor Executor-side adjustments to align outputs with the target game's interfaces. For the three evaluated games, this adaptation required modifying only two lines at the \agentPy\ configuration level, along with an average of 123 lines of game-side changes to support communication between the game and \agentPy.
The Planner, Memory System, and overall agent architecture remain unchanged, enabling scalable deployment across diverse game environments.

\section{Conclusion and Future Work}
\label{sec:conclusion}



This paper presents \agentPy, a Python-based testing tool that operationalizes personality-driven LLM agents for practical game testing. Building on our prior research framework~\cite{MIMIC_YIFEI_ASE_2025}, \agentPy refactors planning, execution, and memory into modular, reusable components and improves deployability over the original JavaScript prototype by decoupling core agent logic from game-specific integration, enabling adaptation to new games through lightweight configuration, prompt updates, and interaction bridges.

However, efficiency of \agentPy remains an important direction for future refinement. In our experiments, each action averaged 12.4 seconds, making \agentPy unsuitable for time-sensitive genres such as First-Person Shooter (FPS) games. The monetary cost was approximately \$0.06 USD per action with code generation and \$0.05 USD without it, so long sessions with thousands of actions can become costly and limit larger-scale adoption.

Looking ahead, we plan to improve \agentPy's efficiency by incorporating fine-tuned local models to reduce latency and cost. These improvements will make personality-driven LLM testing more practical for large-scale, time-sensitive settings. Beyond games, the same modular, personality-driven design can extend to other interactive systems, such as UI testing and human-computer interaction, where behavioural diversity helps uncover edge cases. Together, these contributions position \agentPy as a practical foundation for scalable, behaviourally rich automated testing.



\bibliographystyle{ACM-Reference-Format}

\newpage

\appendix

\section{\agentPy Configuration}
\label{app:configuration}

This section describes the general configuration steps required to use \agentPy. These steps are shared across all game environments and include initializing the agent and selecting personality traits. Once this configuration is completed, the execution workflow of \agentPy remains consistent across different games.

\subsection{Environment Setup}
\agentPy requires Python~3.12. After cloning the repository at \href{https://github.com/Mimic-Persona/MIMIC-Py}{https://github.com/Mimic-Persona/MIMIC-Py}, install the required dependencies by running:

\colorbox{LightGray}{\texttt{pip install -r requirements.txt}}

To simplify deployment and ensure reproducibility, we also provide a preconfigured \href{https://www.virtualbox.org/}{VirtualBox} virtual machine\footnote{Our Virtual Machine: \href{https://drive.google.com/drive/folders/13YUvZ-rPbLjcrUvrajprqPhJnQdfbwKN?usp=sharing}{The Google Drive}}. This VM bypasses most environment setup. Both options are fully documented in the repository README and the following walkthrough.








\subsection{Configuring Parameters}

After cloning our repository, you will find a file named \linebreak \texttt{.env.keep.this} in the root directory. This file serves as a template for creating your own \texttt{.env} configuration file for running \agentPy. The fields that require user modifications are clearly marked in the template and README file and are illustrated in Listing~\ref{listing:1}.

The following parameters must be configured before running \agentPy:

\begin{lstlisting}[
    float=!ht,
    language=Python,
    frame=lines,
    basicstyle=\footnotesize\ttfamily,
    numbers=left,
    numbersep=4pt,
    xleftmargin=0pt,
    aboveskip=0.5em,
    belowskip=0.5em,
    caption={Example of a \texttt{.env} configuration for configuring \agentPy.},
    label={listing:1}
]
#### General Settings ####
## Game settings ##
GAME_SUBJECT=MC

## Agent Personality Settings ##
# Choose from: achievement, adrenaline, aggression, caution,
# completion, curiosity, efficiency, or your custom personalities
PERSONALITY=achievement
# Any name for this agent
AGENT_NAME=achievement1

## Agent Settings ##
# How long do you want the agent to run in minutes
EXP_DURATION=125
# Whether to continue from previous memories with the same AGENT_NAME
IS_CONTINUED=true

## LLM Model Settings ##
# API Keys if needed
OPENAI_API_KEY=sk-proj-...

# Instruction Model
INSTRUCTION_MODEL_NAME=openai/gpt-4o
INSTRUCTION_MODEL_URL=

# Code Model (if needed)
# You have to involve this when running MIMIC-Py in Minecraft or
# other games require Plan-to-Code for interaction
CODE_MODEL_NAME=ollama_chat/XXX:Xb
CODE_MODEL_URL=http://localhost:11434
\end{lstlisting}

\begin{enumerate}
    \item Set \texttt{GAME\_SUBJECT} to the target game environment. The predefined options are \texttt{MC}, \texttt{SPD}, and \texttt{DA}, corresponding to the three games described in Section~\ref{app:game-configuration}.

    \item Set \texttt{PERSONALITY} to the desired personality trait for \agentPy. Supported options include:
    \texttt{achievement}, \texttt{adrenaline}, \texttt{aggression}, \texttt{caution}, \texttt{completion}, \texttt{curiosity}, and \texttt{efficiency}.

    \item Set \texttt{EXP\_DURATION} to specify the execution time limit (in minutes) for a testing session. The default value is 125 minutes.

    \item Set \texttt{IS\_CONTINUED} to control whether the agent resumes from a previous testing session.
    \begin{itemize}
        \item If set to \texttt{true}, \agentPy loads the Memories and Skills associated with the specified \texttt{AGENT\_NAME} and continues execution.
        \item If set to \texttt{false}, \agentPy starts a fresh session and \textbf{deletes} any existing data associated with the same \texttt{AGENT\_NAME}.
        \item We recommend setting this option to \texttt{true}, even for first-time runs.
    \end{itemize}

    \item Configure the instruction model by setting
    \texttt{INSTRUCTION\_MODEL\_NAME} and \texttt{INSTRUCTION\_MODEL\_URL} according to the inference backend in use.

    \begin{enumerate}
        \item \textbf{Using OpenAI GPT models:}
        \begin{itemize}
            \item Set \texttt{OPENAI\_API\_KEY} to a valid OpenAI API key, obtainable from \href{https://platform.openai.com/settings/organization/api-keys}{OpenAI}.
            \item The key typically follows the format \texttt{sk-XXX...} or \texttt{sk-proj-XXX...}.
            \item For example, when using GPT-4o:
        \begin{lstlisting}[float=!ht]
        INSTRUCTION_MODEL_NAME=openai/gpt-4o
        INSTRUCTION_MODEL_URL=
        \end{lstlisting}
        \end{itemize}

        \item \textbf{Using Ollama models:}
        \begin{itemize}
            \item Run the desired model using \href{https://ollama.com/}{Ollama}.
            \item Expose the service via the default endpoint: \linebreak
            \texttt{http://localhost:11434}.
            \item For example:

        \begin{lstlisting}[float=!ht]
        INSTRUCTION_MODEL_NAME=ollama_chat/XXX:Xb
        INSTRUCTION_MODEL_URL=http://localhost:11434
        \end{lstlisting}
        \end{itemize}
    \end{enumerate}

    \item When running \agentPy in environments that require code-based interaction (e.g., Minecraft using Plan-to-Code), configure the code generation model by setting \texttt{CODE\_MODEL\_NAME} and \texttt{CODE\_MODEL\_URL} as shown in (5).

    \item At this stage, the general configuration of \agentPy is complete. After following the game-specific configuration steps in Section~\ref{app:game-configuration}, \agentPy can be deployed to the target game environment by running \texttt{run.py} under the root directory.
\end{enumerate}

\section{Game-Specific Configuration}
\label{app:game-configuration}

This section describes configuration steps specific to each game environment where \agentPy is deployed. These steps are required only to enable \agentPy to interact with a given game and are independent of \agentPy's internal complexity. Instead, they reflect minimal environment-specific requirements imposed by each game's interaction interfaces.

\subsection{Dungeon Adventures}
\label{app:da}
This subsection describes the configuration required to connect \agentPy to the
\textit{Dungeon Adventures} environment used in our evaluation.

\paragraph{\textbf{Launching Dungeon Adventures}}

\begin{enumerate}
    \item Ensure that a Java Development Kit (JDK) is installed on your machine.
    We tested \agentPy with \href{https://www.oracle.com/java/technologies/javase/jdk22-archive-downloads.html}{Java~22}. Users who are using our virtual machine can skip this step.

    \item Start the game by running the main application entry point:

    \colorbox{LightGray}{\texttt{./src/main/java/com/codecool/dungeoncrawl/App.java}}

    \item In the game window, click \texttt{New Adventure}, then select
    \texttt{Start the Game} to launch the game and initialize the server.

    \item Once launched, the terminal will display initialization logs. A shortened example of the output is shown below:
    \begin{lstlisting}[
        float=!h,
        basicstyle=\footnotesize\ttfamily,
        backgroundcolor=\color{LightGray},
        breaklines=true,
        caption={},
        label={listing:da-init}
    ]
    ...
    $$ Agent Mode Enabled!
    \end{lstlisting}
\end{enumerate}

\paragraph{\textbf{Note}}
The following error messages may appear in the console. These messages originate
from the JaCoCo code coverage tool and do \emph{not} affect the functionality of either
\agentPy or the game. They can be safely ignored.
    \begin{lstlisting}[
        float=!h,
        basicstyle=\footnotesize\ttfamily,
        backgroundcolor=\color{LightGray},
        breaklines=true,
        caption={},
        label={listing:da-init}
    ]
    java.net.ConnectException: Connection refused: connect
    ...
    \end{lstlisting}
    
\paragraph{\textbf{Running \agentPy in Dungeon Adventures}}

\begin{enumerate}
    \item Once the game is running and the agent mode is enabled, start \agentPy by running the \texttt{run.py} script. This can be done either by executing the script directly or by running the following command from the project directory:

    \colorbox{LightGray}{\texttt{    python ../run.py}}

    \item After launching \agentPy, the terminal will show initialization logs
    indicating that the agent has successfully connected to the game environment.
    A shortened example is shown below:
    \begin{lstlisting}[
        float=!h,
        basicstyle=\footnotesize\ttfamily,
        backgroundcolor=\color{LightGray},
        breaklines=true,
        caption={},
        label={listing:da-init}
    ]
    ...
    Agent connected to WebSocket server.
    Memory collection initialized.
    ...
    \end{lstlisting}

    \item Once \agentPy connected to the game, press \texttt{B} to start the agent.
\end{enumerate}

\subsection{Shattered Pixel Dungeon}
\label{app:spd}
This subsection describes the configuration required to connect \agentPy to the \textit{Shattered Pixel Dungeon} environment used in our evaluation.

\paragraph{\textbf{Launching Shattered Pixel Dungeon}}

\begin{enumerate}
    \item Ensure that a Java Development Kit (JDK) is installed on your machine.
    We tested \agentPy with \href{https://www.oracle.com/ca-en/java/technologies/downloads/#java21}{Java~21}. Users who are using our virtual machine can skip this step.

    \item Navigate to the \texttt{./MIMIC\_Shattered\_Pixel\_Dungeon} directory:

    \colorbox{LightGray}{\texttt{    cd ./MIMIC\_Shattered\_Pixel\_Dungeon}}

    \item Start the game in debug mode by running:

    \colorbox{LightGray}{\texttt{    ./gradlew desktop:debug}}

    \item In the game window, click \texttt{Enter the Dungeon}, then select \texttt{New Game}. Choose \emph{Warrior} as the character and click \texttt{Start} to begin the game.
    \begin{itemize}
        \item For consistency with our evaluation, only the \textbf{Warrior}
        character is tested in \agentPy.
    \end{itemize}

    \item Once launched, the terminal will display initialization logs. A shortened example of the output is shown below:
        \begin{lstlisting}[
            float=!h,
            basicstyle=\footnotesize\ttfamily,
            backgroundcolor=\color{LightGray},
            breaklines=true,
            caption={},
            label={listing:da-init}
        ]
        > Task :desktop:debug
        [Controllers] added manager for application
        [GAME] @@ You descend to floor 1 of the dungeon.
        ...
        [GAME] $$ Game Server Opened!
        ...
        \end{lstlisting}
\end{enumerate}

\paragraph{\textbf{Note}}
The following error messages may appear in the console. These messages originate
from the JaCoCo code coverage tool and do \emph{not} affect the functionality of either \agentPy or the game. They can be safely ignored.

    \begin{lstlisting}[
        float=!h,
        basicstyle=\footnotesize\ttfamily,
        backgroundcolor=\color{LightGray},
        breaklines=true,
        caption={},
        label={listing:da-init}
    ]
    java.net.ConnectException: Connection refused: connect
    ...
    \end{lstlisting}
\paragraph{\textbf{Running \agentPy in Shattered Pixel Dungeon}}

\begin{enumerate}
    \item While in the game, press the \texttt{``B''} key. A pop-up window will appear, prompting you to enter a command. At the same time, the console should display:
    \begin{lstlisting}[
        float=!h,
        basicstyle=\footnotesize\ttfamily,
        backgroundcolor=\color{LightGray},
        breaklines=true,
        caption={},
        label={listing:da-init}
    ]
    [GAME] $$ MIMIC Mode Started with XXX milliseconds!
    \end{lstlisting}

    \item Once the MIMIC mode is active, start \agentPy by running the \texttt{run.py} script. This can be done either by executing the script directly or by running the following command from the \texttt{./MIMIC\_Shattered\_Pixel\_Dungeon} directory:

    \colorbox{LightGray}{\texttt{    python ../run.py}}

    \item After launching \agentPy, the terminal will show initialization logs indicating that the agent has successfully connected to the game environment.
    A shortened example is shown below:
    \begin{lstlisting}[
        float=!h,
        basicstyle=\footnotesize\ttfamily,
        backgroundcolor=\color{LightGray},
        breaklines=true,
        caption={},
        label={listing:da-init}
    ]
        ...
        Agent connected to WebSocket server.
        Memory collection initialized.
        ...
    \end{lstlisting}

    \item Finally, return to the game pop-up window, enter the command \texttt{1}, and click the \texttt{Set} button to start \agentPy.
\end{enumerate}

\subsection{Minecraft}
\label{app:minecraft}

This subsection describes the configuration required to connect \agentPy to the \textit{Minecraft} environment used in our evaluation.

\paragraph{\textbf{Minecraft Environment Setup (External)}}
To run \agentPy in Minecraft, users must set up a compatible Minecraft environment by installing the required mods and data packs and starting a local LAN server. To simplify deployment and ensure reproducibility, we also provide a preconfigured \href{https://www.virtualbox.org/}{VirtualBox} virtual machine\footnote{Our Virtual Machine: \href{https://drive.google.com/drive/folders/13YUvZ-rPbLjcrUvrajprqPhJnQdfbwKN?usp=sharing}{The Google Drive}} that bypasses manual environment setup. Both options are fully documented in the README for Minecraft\footnote{Our README for \agentPy in MC: \href{https://github.com/Mimic-Persona/MIMIC-Py/tree/main/MIMIC_Minecraft}{MC README}}.

\begin{figure}[!h]
        \centering
        \includegraphics[width=\linewidth]{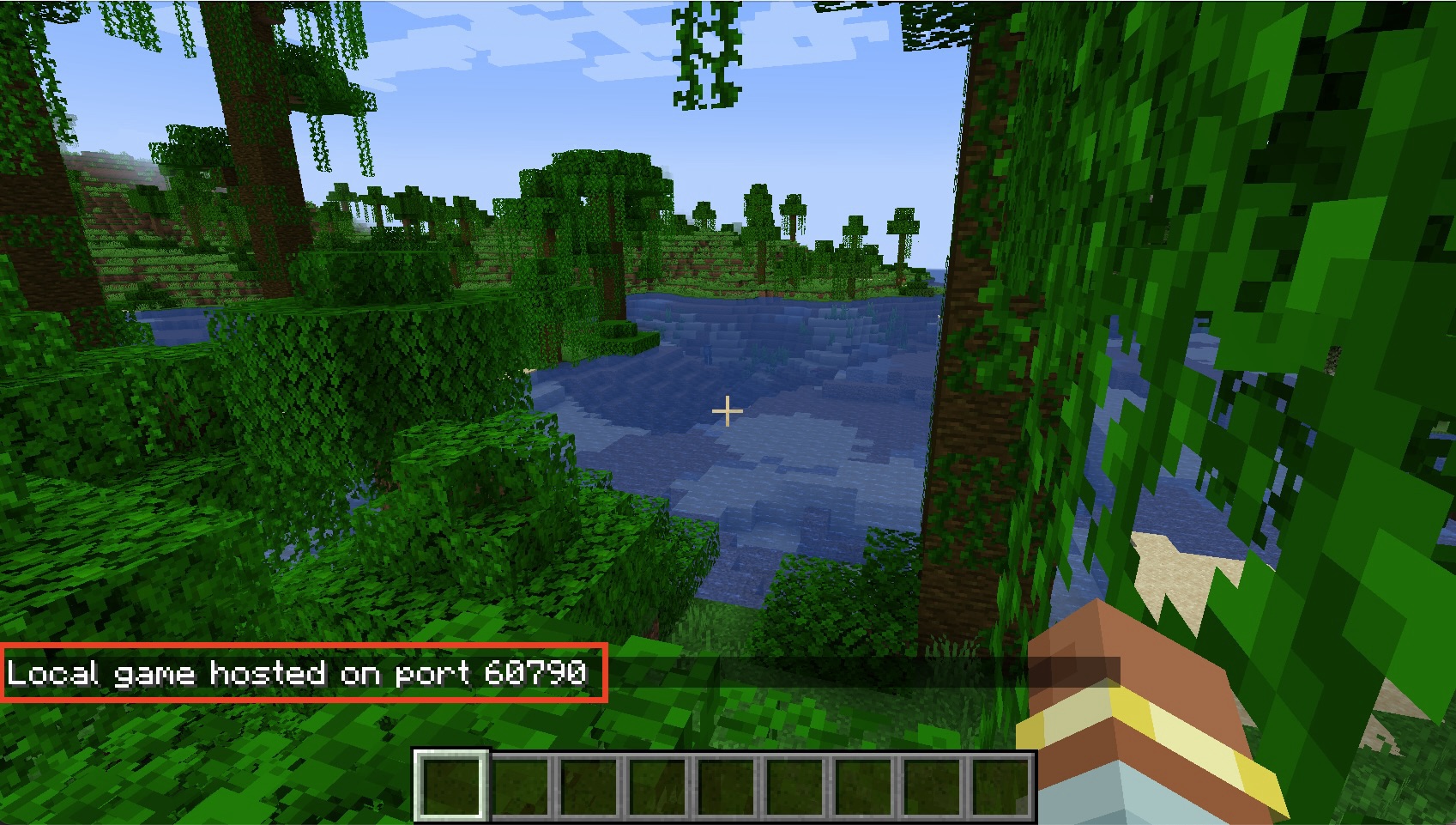}
        \caption{Confirmation message showing the LAN server port number.}
        \label{fig:mc10}
    \end{figure}

Once the Minecraft world is launched and opened to LAN, the game will display a port number indicating the active local server (see \figurename~\ref{fig:mc10}). Record this port number and set it as the value of \texttt{MC\_PORT} in the \texttt{.env} file located at the root directory. This port is used by \agentPy to connect to the Minecraft server. Detailed instructions for configuring the \texttt{.env} file for Minecraft-specific parameters are provided later.

If users choose to use the virtual machine we provided, they can skip the next step to the \texttt{Configuring Parameters} step.

\paragraph{\textbf{Installing Node.js and Dependencies}}
\agentPy interacts with Minecraft via a third-party API library~\cite{Mineflayer}, which requires Node.js. The following steps describe how to install Node.js and the required dependencies for enabling interaction between \agentPy and the Minecraft environment.

\begin{enumerate}
    \item Ensure that \texttt{Node.js} is installed on your machine.  
    \textbf{Important:} Only \texttt{Node.js LTS} versions (e.g., 22.x or 24.x) are supported.  
    Newer Node versions (e.g., 25) are incompatible with \texttt{mineflayer} due to deprecated APIs.  
    Node.js can be downloaded from the \href{https://nodejs.org/en/download}{official website}.

    \item Open a new terminal and navigate to the \texttt{./MIMIC\_Minecraft} directory:

    \colorbox{LightGray}{\texttt{    cd ./MIMIC\_Minecraft}}

    \item Install the required Node.js dependencies:

    \colorbox{LightGray}{\texttt{    npm install chromadb@1.10.5}}
    
    \colorbox{LightGray}{\texttt{    npm install}}

    \item Install the modified \texttt{mineflayer-collectblock} plugin:

    \colorbox{LightGray}{\texttt{    cd ./mc\_env/mineflayer/mineflayer-collectblock}}

    \colorbox{LightGray}{\texttt{    npm install mineflayer-collectblock}}
    
    \item Navigate to the directory containing the modified Mineflayer library and install its dependencies:
    
    \colorbox{LightGray}{\texttt{    cd ./mc\_env/mineflayer}}
    
    \colorbox{LightGray}{\texttt{    npm install}}

    \item If compatibility issues arise during execution, remove all \texttt{node\_modules} directories in the paths above and reinstall dependencies.

    \item If a \texttt{MODULE\_NOT\_FOUND} error related to mineflayer-collectblock occurs, please try to reinstall the `mineflayer` package after making sure the `mineflayer-collectblock` is correctly built from step 4.
\end{enumerate}

\begin{lstlisting}[
    float=!ht,
    language=Python,
    frame=lines,
    framesep=2mm,
    backgroundcolor=\color{LightGray},
    basicstyle=\footnotesize\ttfamily,
    numbers=left,
    numbersep=4pt,
    showstringspaces=false,
    breaklines=true,
    columns=fullflexible,
    keepspaces=true,
    caption={Example of a \texttt{.env} configuration for running \agentPy in Minecraft.},
    label={listing:2}
]
#### Settings for Minecraft Only ####
## Minecraft Host and Port ##
MC_HOST=localhost
MC_PORT=60790

## Minecraft Task Settings ##
# Choose from:
# combat_1_cave_spider, combat_1_skeleton, combat_1_spider,
# cook_1_meat, harvest_1_diamond, harvest_1_sugar,
# sheer_1_sheep, and survive_for_1_day (or your custom tasks).
MC_TASK=shear_1_sheep
MC_TASK_ID=0 # Any number
MC_MONSTER_TYPE=cave_spider # Type of monster for combat tasks
\end{lstlisting}

\paragraph{\textbf{Configuring Parameters}}
An example configuration for this step can be found in the same \texttt{.env.keep.this} file located in the root directory, as mentioned earlier. In this subsection, we focus on parameters specific to the Minecraft environment. These parameters are highlighted in Listing~\ref{listing:2}.

\begin{enumerate}
    \item Set \texttt{MC\_PORT} to the port number of the Minecraft LAN server started in the previous step.

    \item Set \texttt{MC\_TASK} to specify the task that \agentPy should perform in Minecraft. The following predefined tasks are supported:
    \begin{itemize}
        \item \texttt{combat\_1\_cave\_spider}
        \item \texttt{combat\_1\_skeleton}
        \item \texttt{combat\_1\_spider}
        \item \texttt{cook\_1\_meat}
        \item \texttt{harvest\_1\_diamond}
        \item \texttt{harvest\_1\_sugar}
        \item \texttt{sheer\_1\_sheep}
        \item \texttt{survive\_for\_1\_day}
    \end{itemize}
    \noindent
    \textbf{Note:} Tasks outside this list do not have predefined task descriptions. Users may still specify arbitrary task names, but such tasks will not be associated with structured task descriptions for the Planner.

    \item Set \texttt{MC\_MONSTER\_TYPE} to specify the monster type used in combat-related tasks. Tested options include:
    \begin{itemize}
        \item \texttt{cave\_spider}
        \item \texttt{skeleton}
        \item \texttt{spider}
    \end{itemize}
    \noindent
    \textbf{Note:} This parameter is required only when a combat-related task is selected. The specified monster type will be spawned automatically after one in-game day (approximately 20 minutes) for the agent to engage. Other monster types supported by Minecraft may also be used.
\end{enumerate}

\paragraph{\textbf{Running \agentPy in Minecraft}}
\begin{enumerate}
    \item After completing all configurations in the \texttt{.env} file, start \agentPy by running the \texttt{run.py} script. This can be done either by executing the script directly or by running the following command from the \texttt{./MIMIC\_Minecraft} directory:

    \colorbox{LightGray}{\texttt{    python ../run.py}}

    \item Once launched, the terminal will display initialization logs. A shortened example of the output is shown below:
        \begin{lstlisting}[
        float=!h,
        basicstyle=\footnotesize\ttfamily,
        backgroundcolor=\color{LightGray},
        showstringspaces=false,
        breaklines=true,
        columns=fullflexible,
        keepspaces=true,
        label={listing:4}
    ]
    [INFO] File written successfully
    ...
    Subprocess mineflayer started with PID XXX.
    mineflayer ready line: Server started on port 3000
    ...
    INFO:Socket: Connected to localhost
    INFO:mineflayer: Python bridge connected
    ...
    Memory System initialized
    Skill System initialized
    \end{lstlisting}

    \item At the same time, status messages confirming that \agentPy has successfully connected will appear in the Minecraft chat window, as illustrated in \figurename~\ref{fig:mc11}.
    \begin{figure}[!h]
        \centering
        \includegraphics[width=\linewidth]{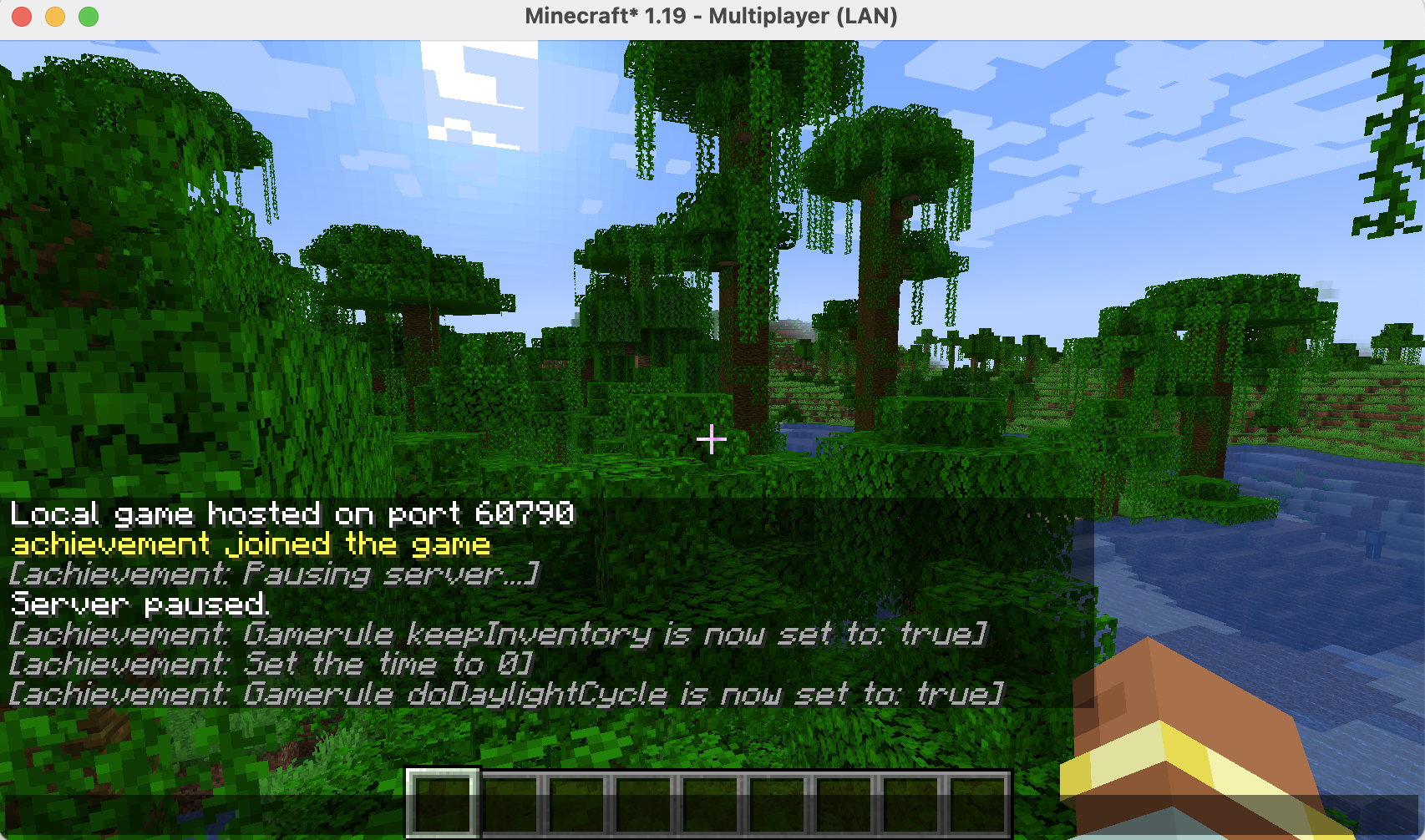}
        \caption{Minecraft chat window showing successful connection to \agentPy.}
        \label{fig:mc11}
    \end{figure}

    \item Press the \texttt{“T”} key to open the in-game chat window, then send message “b” to start the \agentPy agent.
\end{enumerate}






\section{Deploying \agentPy to New Game Environments}
\label{app:deployment-new-games}

This section outlines the minimal steps required to deploy \agentPy to a previously unsupported game environment. The design intentionally isolates game-specific engineering effort from the core agent logic, allowing new environments to be integrated without modifying the Planner, Memory System, or Skill System.

\subsection{Game State Representation}

To support decision making, \agentPy requires a structured representation of the current game state.

\begin{itemize}
    \item Users must define a game-state abstraction that captures all information relevant to planning and execution, such as player status, environment context, inventory, and nearby entities.
    \item This representation should be extractable from the game environment at runtime and serializable for communication with the agent.
\end{itemize}

Once defined, users implement environment-side logic to extract and transmit the game state whenever the agent requests it.

\subsection{Prompt Adaptation (Repository-Guided)}

\agentPy relies on prompt templates to ground planning and reasoning in game-specific mechanics. Rather than detailing prompt engineering in this paper, we abstract this step as follows:

\begin{itemize}
    \item Prompt templates are organized in the repository under \texttt{./prompts/template/}\footnote{Our prompt templates: \href{https://github.com/Mimic-Persona/MIMIC-Py/tree/main/prompts/template}{The Templates}}.
    \item To deploy \agentPy to a new game, users copy the provided templates into a new game-specific folder and update placeholders (e.g., game state descriptions, task semantics, and game rules).
    \item After modifying the templates for the target game, the agent can be executed without further changes to the planning pipeline.
\end{itemize}

Detailed prompt adaptation guidelines and examples are provided in the project repository.

\subsection{Game Interaction APIs}

\agentPy interacts with game environments via a lightweight socket-based bridge that defines a small set of interaction APIs that constitute the contract between \agentPy and a game environment.
By default, communication is established over a socket listening on \texttt{localhost:1111}, which can be reconfigured via the \texttt{.env} file.

\paragraph{\textbf{Environment Interaction APIs}}
\agentPy communicates with game environments through a socket-based bridge that exposes four core APIs. Each API has a well-defined role and invocation context during execution.

\begin{enumerate}
    \item \texttt{get\_command()}:
    Blocks until a socket message with \texttt{msgType=``command''} is received from the game environment, then returns the corresponding command string. This API is invoked \emph{once at startup}, and \agentPy blocks on this call until the environment sends the start signal \texttt{``b''}, synchronizing agent initialization with the environment.

    \item \texttt{get\_status()}:
    Sends a socket message with \texttt{``GetStatus''} and blocks until a response with \texttt{msgType=``status''} is received. The returned value is a serialized snapshot of the current game state. This API is invoked at the \emph{beginning of each interaction iteration} and \emph{after each action execution} to obtain updated state information.

    \item \texttt{act\_and\_feedback(plan)}:
    Sends an action plan to the game environment using a socket message prefixed with \texttt{``ACTION:''}, where \texttt{plan} is the Planner output serialized as JSON. The call blocks until the environment completes execution and responds with a message of \texttt{msgType=``feedback''}, which contains (a) \texttt{logs}: execution logs generated by the environment, (b) \texttt{errors}: error messages encountered during execution. The API returns structured feedback objects derived from these fields. This API is used \emph{only} when the Plan-to-Parameters interaction mechanism is enabled, where the environment interprets and executes structured action plans.

    \item \texttt{run\_and\_feedback(code, programs, timeout, executor)}:
    Executes generated code within the game environment using user-provided, game-specific execution support. The generated code runs with access to \texttt{programs} as helper functions and under an enforced runtime \texttt{timeout}. Execution is delegated to an \texttt{executor} function supplied by the user, which is responsible for executing the code inside the target environment and returning
    (i) a serialized game-state observation and
    (ii) execution metadata, including timeout status, log messages, and error messages.
    The call blocks until execution completes and returns both the observation and the execution metadata.
    
    This API is invoked \emph{only} when the \emph{Plan-to-Code} interaction mechanism is enabled. A reference executor implementation is provided for Minecraft. For other game environments, users must implement a compatible executor function and register it via configuration in \texttt{run.py}.
\end{enumerate}

\paragraph{\textbf{Implementation Requirements for New Game Deployment}}

When deploying \agentPy to a new game, users must implement the corresponding environment-side interfaces for the selected interaction mode. \agentPy supports two mutually exclusive interaction modes for connecting the agent to a game environment:

\begin{itemize}
    \item \textbf{Plan-to-Parameters:}  
    The agent generates structured action plans (e.g., JSON), which the game environment interprets and executes via \texttt{act\_and\_feedback}. In this mode, users must implement environment-side handlers that map these plans to concrete game actions.

    \item \textbf{Plan-to-Code:}  
    The agent generates executable code snippets that interact directly with the game environment through helper functions and are executed via \texttt{run\_and\_feedback}.
    Unlike Plan-to-Parameters, this mode requires users to implement a \emph{game-specific code execution layer}, exposed as an \texttt{executor} function, which safely executes generated code inside the target environment and returns structured execution feedback (e.g., timeout status, logs, and errors) to \agentPy.
    A reference implementation of such an executor is provided in \texttt{MineEnv.py}\footnote{Example of code executor for game interaction: \href{https://github.com/Mimic-Persona/MIMIC-Py/blob/main/MIMIC_Minecraft/mc_env/MineEnv.py}{MineEnv.py}}.
    
    In addition, users must supply a small set of initial \emph{Basic Skills}, consisting of example code snippets and accompanying textual descriptions that demonstrate how the game's APIs can be invoked.
    These Basic Skills serve two purposes: (i) they guide the Plan-to-Code Translator in synthesizing valid code, and (ii) they act as reusable helper functions for incrementally constructing more complex Skills.
    Representative examples are available in the repository\footnote{Examples of basic Skills: \href{https://github.com/Mimic-Persona/MIMIC-Py/tree/main/skill_system/skill_library/MC/basic_skills}{The Basic Skills for MC}}.
\end{itemize}

\subsection{Registering New Environments}

After implementing the required game interaction APIs, register the new environment by setting \texttt{GAME\_SUBJECT} to the corresponding identifier and configuring the \texttt{IS\_PLAN\_TO\_CODE} flag in the \texttt{.env} file to select the appropriate interaction mode.

In summary, deploying \agentPy to a new game environment involves three main steps: (1) defining an abstract game state representation, (2) adapting prompt templates to reflect game-specific entities and rules, and (3) implementing a lightweight interaction bridge between \agentPy and the game. These steps localize all environment-specific engineering effort, allowing the core agent architecture to remain unchanged across different testing environments.

\end{document}